\documentclass[12pt]{iopart}
\usepackage{graphicx,iopams}

\begin{document}

\title{Analyze This! A cosmological constraint package for cmbeasy}
\author{Michael Doran\dag and Christian M M{\"{u}}ller\ddag }

\address{\dag Department of Physics and Astronomy, Dartmouth College, 6127 Wilder Laboratory, Hanover, New Hampshire 03755, USA}

\address{\ddag Institut f{\"{u}}r Theoretische Physik, Philosophenweg 16, 69120 Heidelberg, Germany}

\ead{\mailto{Michael.Doran@Dartmouth.edu}, \mailto{C.M.Mueller@thphys.uni-heidelberg.de}}

%\url{www.cmbeasy.org}

\date{May 4, 2004}

\newcommand{\postd}{\pi(\theta|X)}
\newcommand{\covM}{\bi{S}}

\begin{abstract}
  We introduce a Markov Chain Monte Carlo simulation and data analysis
  package that extends the \textsc{Cmbeasy} software.  We have taken
  special care in implementing an adaptive step algorithm for the
  Markov Chain Monte Carlo in order to improve convergence.
  Data analysis routines are provided which allow to
  test models of the Universe against measurements of the cosmic
  microwave background, supernovae Ia and large scale structure. We present 
  constraints on cosmological parameters derived from these measurements for a $\Lambda$CDM cosmology
and 
  discuss the impact of the different observational data sets on the parameters. 
  The package is publicly available as  part of the {\sc Cmbeasy} software at www.cmbeasy.org.
\end{abstract}

%\preprint{HD--THEP--03--57}

\newcommand{\class}[1]{{\tt #1}}
\maketitle

%%%%%%%%%%%%%%%%%%%%%%%%%%%%%%%%%%%%%%%%%%%%%%%%%%%%%%%%%%%%%%%%%%%%%%%%%%%%%%%%%%%%%%%%%%%%%%%%%%%

\section{Introduction}
The wealth of recent  precision measurements in cosmology 
\cite{Bennett:2003bz,Hinshaw:2003ex,Kogut:2003et,Kuo:2002ua,Readhead:2004gy,Dickinson:2004yr,Riess:2004nr,Tonry:2003zg,Knop:2003iy,Barris:2003dq,Percival:2001hw,Verde:2002,Peacock:2001gs,Tegmark:2003ud,Tegmark:2003uf} 
places stringent constraints on any model of the Universe. Typically, such a model is given in terms of a number
of cosmological parameters. Numerical tools, such as {\sc Cmbfast} \cite{Seljak:1996is}, 
\textsc{Camb}  \cite{Lewis:1999bs} and \mbox{\sc Cmbeasy}  \cite{Doran:2003sy}, permit to calculate 
the prediction of a given model  for the observational data. While these tools are  comparatively fast, 
scanning the parameter space for the most likely model and confidence regions can become a matter of 
time and computing power. 

The cost of evaluating models on a n-dimensional grid in parameter space 
increases exponentially with the number of parameters. 
In contrast, the Markov Chain Monte Carlo (MCMC) method scales 
roughly linearly with the number of parameters \cite{Christensen:2000ji,Christensen:2001gj,Lewis:2002ah}.  The MCMC
method has already been used to constrain  various models \cite{Knox:2001fz,Kosowsky:2002zt,Verde:2003ey,Caldwell:2003hz}.
A popular tool for setting up MCMC simulations is the \textsc{Cosmo-Mc} package \cite{Lewis:2002ah} 
for the  \textsc{Camb} code, an improved proposal distribution for the local Metropolis algorithm 
has been proposed in \cite{Slosar:2003da}.

\begin{figure}
\begin{center}
\includegraphics[scale=5]{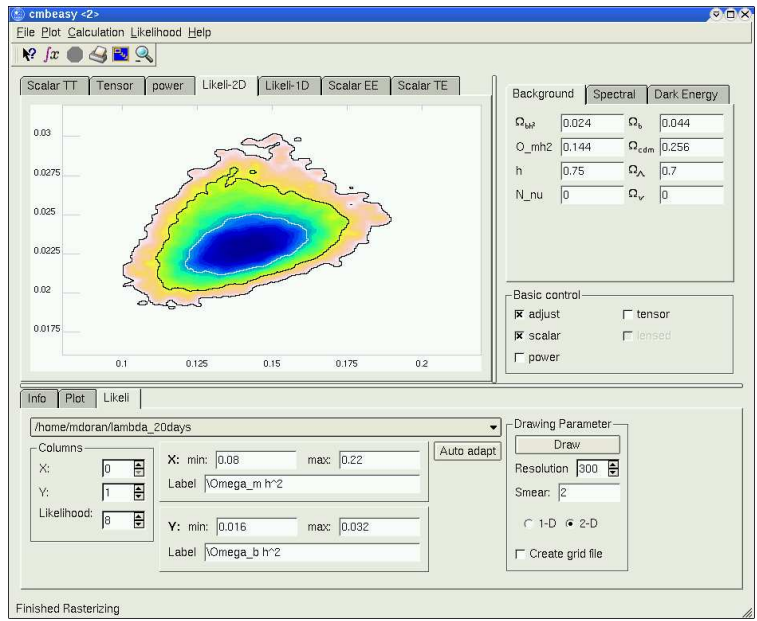}
\caption{\label{fig::gui}The graphical user interface of {\sc Cmbeasy}. It can be used to 
marginalize, visualize and print the one and two dimensional likelihoods from the MCMC chains. 
Shown is the marginalized likelihood in the $\Omega_m h^2 - \Omega_b h^2$ plane of a $\Lambda$CDM model. } 
\end{center}
\end{figure}

In this paper, we  introduce the \class{AnalyzeThis} package\footnote{It is part of the cmbeasy v2.0 release.}
 for {\sc Cmbeasy}. It includes a
parallel MCMC driver, as well as
routines 
to calculate the likelihood of a model with respect to various data sets. 
We took special care in designing a step-proposal strategy that leads to fast
convergence and mixing of the chains. This strategy is applied during the early
stages of the simulation. As soon as the likelihood contour has been roughly
explored, the adaptive step proposal freezes in. 
This ensures that the MCMC results are not contaminated by
the adaptive steps. At the same time, the automated step optimization 
considerably improves performance and is rather convenient. 
The raw data files can be processed from within a graphical user interface (gui).
Using the gui, one can marginalize, visualize and print  one and two
dimensional likelihood surfaces (see figure \ref{fig::gui}).

The plan of this paper is as follows: we describe the MCMC method and our implementation
in section \ref{sec::simulation}. A brief introduction to the software is given in section \ref{sec::user}.
We analyze the constraints on cosmological parameters from observational data sets in section  \ref{sec::constraints}.
In section \ref{sec::conclusion}  we present our conclusions, while the format of the MCMC
data files is defined in \ref{appendix::format}.

%%%%%%%%%%%%%%%%%%%%%%%%%%%%%%%%%%%%%%%%%%%%%%%%%%%%%%%%%%%%%%%%%%%%%%%%%%%%%%%%%%%%%%%%%%%%%%%%%%%%
\section{Markov Chain Monte Carlo simulation}\label{sec::simulation}
In the following, we will review the basic ideas of Markov Chain
Monte Carlo simulation  \cite{Gilks,Gamerman,Neal} and describe our implementation. 

Suppose that
$\theta$ denotes a vector of all model parameters, $X$ represents some
observed data and $L(X|\theta)$ is the likelihood of observing $X$
given parameters $\theta$.
Specifying a prior distribution $P(\theta)$ for the parameters, Bayes' theorem
yields the posterior distribution  $\postd$ of $\theta$ given the observed data $X$:
\begin{equation}
\postd = \frac{P(\theta)L(X|\theta)}{\int P(\theta)L(X|\theta)d\theta}.
\end{equation}
Using  $\postd$, one can
compute
expectation values 
\begin{equation}
E[f(\theta)|X]= \frac{\int f(\theta) \postd d\theta}{\int \postd d\theta},
\end{equation}
as well as confidence levels. 

The idea of the MCMC method is to directly draw samples from the posterior $\postd$.
The statistical properties of $\postd$ may then be estimated using this sample.\footnote{For instance, one can
infer the mean of
parameter $\theta^{(i)}$ from the sample of parameter values $\theta_1,...,\theta_n$ (the ergodic average):
\begin{equation}
E[\theta^{(i)}]=\frac{1}{n} \sum_{j=1}^{n} \theta_j^{(i)}.
\end{equation} 
}
To accomplish the sampling of $\postd$ one uses a Markov Chain, which  is a stochastic process
$\{\theta_0,\theta_1,...,\theta_n \}$ where $\theta_n$ only depends on $\theta_{n-1}$.
The idea is to choose the next 
point in the chain based on the previous point such that $\postd$ becomes the
stationary distribution of the chain
\begin{equation}
\mbox{Dist}\{\theta_0,...,\theta_n\} \rightarrow \postd \mbox{ as } n\rightarrow \infty.
\end{equation}
There are several methods to accomplish this. We will concentrate on the Metropolis algorithm \cite{Metropolis:am} 
and its implementation in {\sc Cmbeasy}.
%%%%%%%%%%%%%%%%%%%%%%%%%%%%%%%%%%%%%%%%%%%%%%%%%%%%%%%%%%%%%%%%%%%%%%%%%%%%%%%%%%%%%%%%%%%555

\subsection{The Metropolis Algorithm}\label{sec::metropolis}
\begin{figure}[!t]
\begin{center}
\includegraphics[angle=0,scale=0.5]{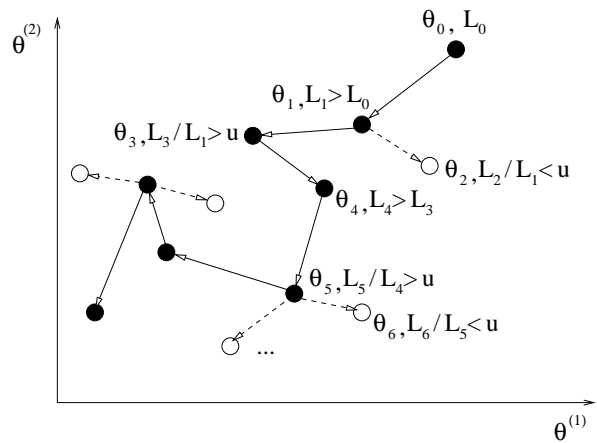}
\caption{Illustrating the Metropolis algorithm for two parameters.
 Filled circles represent points belonging to the chain,
empty circles are proposed but rejected points not belonging to the chain. In this
example, the chain would be $[\theta_0, \theta_1, \theta_1,\theta_3,\theta_4, \dots]$.}
\label{fig::randwalk}
\end{center}
\end{figure}

The algorithm is defined as follows (for an illustration see Fig.~\ref{fig::randwalk}):
\begin{enumerate}
\item Choose starting parameter vector $\theta_0$.
\item Compute the likelihood $L_0(X|\theta_0)$ of observing the experimental data given the parameters $\theta_0$.
\item Obtain a new parameter vector by sampling from  a ``proposal distribution'' $q(\theta_{i-1},\theta_{i})$ (see 
section \ref{sec::adaptive}). \label{choose_point}
\item Compute the likelihood $L_i(X|\theta_i)$.
\item If $L_i > L_{i-1}$ then save $\theta_i$ as new point in the chain (``take the step'') and go to (\ref{choose_point}). \label{take_step}
\item If $L_i < L_{i-1}$ then generate a random variable $u$ from  $[0,1]$. If $u < L_i/L_{i-1}$ take the step as in 
(\ref{take_step}). If $u > L_i/L_{i-1}$ then reject $\theta_i$, save $\theta_{i-1}$ as new point in the chain and go to 
(\ref{choose_point}).
\end{enumerate}
%%%%%%%%%%%%%%%%%%%%%%%%%%%%%%%%%%%%%%%%%%%%%%%%%%%%%%%%%%%%%%%%%%%%%%%%%%%%%%%%%%%%%%%%%%%%%%%%%%%%%

%%%%%%%%%%%%%%%%%%%%%%%%%%%%%%%%%%%%%%%%%%%%%%%%%%%%%%%%%%%%%%%%%%%%%%%%%%%%%%%%%%%%%%%%%%%%%%%%%%%%%%
This algorithm assumes flat priors $P(\theta)$ and a
symmetric proposal distribution
$q(\theta_{i-1},\theta_i)=q(\theta_{i},\theta_{i-1})$. Note that we
assign likelihood zero to any parameter set that has at least one
point outside its prior. In this version of the Metropolis algorithm,
all parameters change with every step, a strategy called global Metropolis.\footnote[1]{
One may also change one parameter (or a subset of parameters) at a
time. Such a  local Metropolis algorithm is implemented in  \textsc{Cosmo-Mc}\cite{Lewis:2002ah}.}

%%%%%%%%%%%%%%%%%%%%%%%%%%%%%%%%%%%%%%%%%%%%%%%%%%%%%%%%%%%%%%%%%%%%%%%%%%%%%%%%%%%%%%%%%%%%%%%%%%%%%%%%%%%%%%

\subsection{Convergence Testing}\label{sec::convergence}
At the beginning, the chain migrates from its random
starting point to regions of higher likelihood. Points during
this ``burn-in'' do not constitute a sample from $\postd$ and should
be eliminated. In principle,
it may be difficult to tell from a single chain if it has converged
towards the underlying $\postd$. In MCMC, one therefore uses several
chains with random starting points and monitors mixing and
convergence.

The convergence test of Gelman and Rubin
\cite{Gelman_Rubin} monitors the variance of a parameter between the chains.
To be precise, consider
using the last $n$ points of each of $m$ chains for the test.  Let
$\psi_{ij}$ label one entry of the parameter vector $\theta$ at point
$j=1,\dots,n$ in chain $i$ with $\overline{ \psi}_i$ the mean for
chain $i$ and $\overline \psi$ the mean of all chains. The variance
between chains $B$ and the within-chain variance $W$ are then given by
\begin{eqnarray}
B&=& \frac{n}{m-1}\sum_{i=1}^{m}(\overline \psi_i - \overline \psi)^2, \\
W&=&\frac{1}{m}\sum_{i=1}^{m}s_i^2, \mbox{ where }  s_i^2=\frac{1}{n-1} \sum_{j=1}^n(\psi_{ij}-\overline\psi_i)^2,
\end{eqnarray}
and the quantity
\begin{equation}
 R =\frac{\frac{n-1}{n}W+\frac{1}{n}B}{W}
\end{equation}
should converge to one.\footnote{In a realistic situation, the numerator is an overestimate whereas the denominator
is an underestimate of the variance of the stationary distribution of $\psi$.}
A value of $R < 1.2$ for all parameters indicates that the chain is sampling from $\postd$.\footnote{Indicating that the step sizes
and directions will freeze in in our implementation.} From this point onwards
one may use the chain points for parameter estimates. 

When do we have enough points for parameter estimation? This question is not easy to answer,  since it is depends on the model, the  used data sets and the desired accuracy
how many chain points are needed for a  robust estimate of parameters. Therefore,
 in our implementation
the MCMC simulation runs indefinetly. However, any ``breaking-criterion'' may be implemented easily, and the chains may be monitored with external programs during the run.

%%%%%%%%%%%%%%%%%%%%%%%%%%%%%%%%%%%%%%%%%%%%%%%%%%%%%%%%%%%%%%%%%%%%%%%%%%%%%%%%%%%%%%%%%%%%%%%%%%%
\subsection{Adaptive Step Size Gaussian Sampler}\label{sec::adaptive}
\begin{figure}[!t]
\begin{center}
\includegraphics[angle=0,scale=0.55]{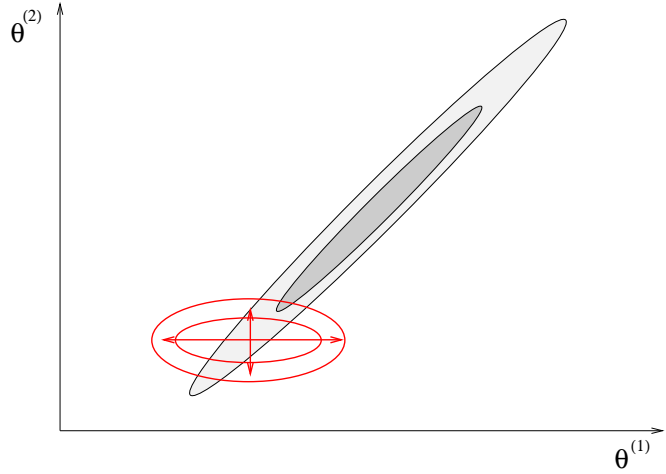}
\caption{Illustrating the naive Gaussian sampler with fixed step sizes for two parameters.
 The (unknown) true likelihood surface is shown in black, the proposal distribution with arrows in red.
This proposal distribution does not 
does not take into account the 
 degeneracy among the parameters $\theta^{(1)}$ and $\theta^{(2)}$, leading to slow mixing.}
\label{fig::likelihood}
\end{center}
\end{figure}

The number of steps needed for good convergence and mixing depends
strongly on the step proposal distribution. If the proposed steps are too
large, the algorithm will frequently reject steps, giving slow
convergence of the chain. If, on the other hand, the proposed steps
are too small, it will take a long time for the chain to explore the
likelihood surface, resulting in slow mixing.  In the ideal case the
proposal distribution should be as close to the posterior distribution $\postd$ as possible -- which
unfortunately is not known a priori.
While a simple Gaussian proposal distribution with step sizes $\sigma_k$ is
sufficient, it is not optimal in terms of computing costs if cosmological parameters
are degenerate.  A naive
Gaussian sampler would move slowly along
the degeneracy direction, unaware of any degeneracy (see figure \ref{fig::likelihood}).

%%%%%%%%%%%%%%%%%%%%%%%%%%%%%%%%%%%%%%%%%%%%%%%%%%%%%%%%%%%%%%%%%%%%%%%%%%%%%%%%%%%%%%%%%%%%%%%%%%%%%

%%%%%%%%%%%%%%%%%%%%%%%%%%%%%%%%%%%%%%%%%%%%%%%%%%%%%%%%%%%%%%%%%%%%%%%%%%%%%%%%%%%%%%%%%%%%%%%%%%%%%%
Instead of using a naive Gaussian proposal distribution, we sample from a multivariate
Gaussian distribution with covariance matrix estimated from the previous points in the chains. 
By taking into account the covariances
among the parameters, we effectively approximate the likelihood contour in extent and orientation --  the
Gaussian samples are taken along the principal axis of the likelihood contour.
Denoting  $\theta_{ i-1}-\theta_i=\bi{u}_i$ for notational convenience, the proposal distribution 
we use for the steps is
\begin{equation}\label{eqn::gauss}
q(\theta_{i-1},\theta_i) \sim N \exp\left[-\frac{1}{2}\bi{u}_i^T\covM^{-1} \bi{u}_i\right],
\end{equation}
%\mbox{with } N&=& \left(2\pi\right)^{-K/2} (\mbox{det}\, \covM)^{-1/2}.
where $N=\left(2\pi\right)^{-K/2} (\mbox{det}\, \covM)^{-1/2}$ and  $\bi{S}$ is the 
covariance matrix 
\begin{equation}
\covM = \left( \begin{array}{cccc}
        \sigma_1^2 & \rho_{12} & \dots & \rho_{1K} \\
        \rho_{21} & \sigma_2^2 & \dots & \rho_{2K} \\
         \vdots &          & \ddots & \vdots  \\
        \rho_{K1} &\dots & \rho_{K-1 K} & \sigma_K^2
        \end{array}\right).
\end{equation}
The sampling is most easily performed by diagonalizing
the covariance matrix
\begin{equation}
\bi{T}^T \covM \, \bi{T}=\bi{D}  \Longleftrightarrow \bi{T}^T 
\covM^{-1} \bi{T}=\bi{D}^{-1},
\end{equation}
where $\bi{T}$ is an orthogonal matrix. Using this, equation (\ref{eqn::gauss}) becomes
\begin{eqnarray}
&&q(\theta_{i-1},\theta_{i}) \sim N \exp\left[-\bi{u}_i^T \bi{T} \bi{T}^T\covM^{-1}
  \,\bi{T}    \bi{T}^T \bi{u}_i\right] \\ 
&&=  N \exp\left[-\frac{1}{2}\bi{v}_i^T \bi{D}^{-1} \bi{v}_i\right],
\end{eqnarray}
where $\bi{v}_i \equiv \bi{T}^T \bi{u}_i$.
Thus, the procedure for obtaining a sample $\bi{u}_i$ is as follows:
\begin{enumerate}
\item Find the eigenvalues $\tilde \sigma_j^2$ and eigenvectors  of $\covM$. Construct the
transformation matrix $\bi{T}$ from the eigenvectors. 
\item Draw Gaussian samples with variances $\tilde \sigma_j^2$, thereby 
obtaining the vector $\bi{v}_i$.
\item Then $\bi{u}_i=\bi{T} \bi{v}_i$ is the desired 
sample from the multivariate Gaussian with covariance matrix $\covM$.
\end{enumerate}

\begin{figure}[!t]
\begin{center}
\includegraphics[angle=0,scale=0.42]{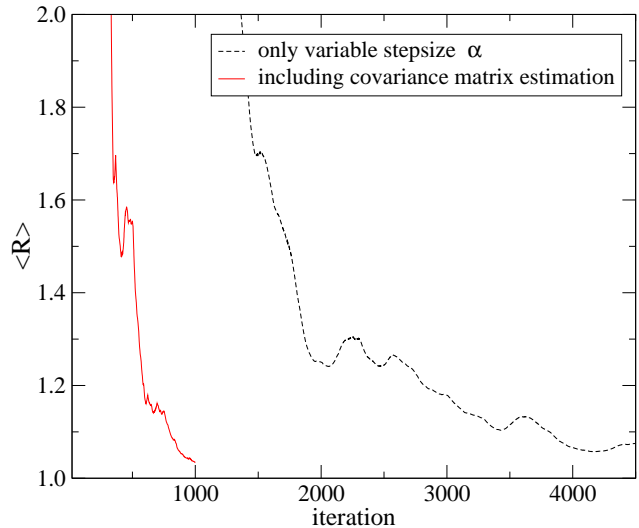}
\caption{Convergence properties using different proposal distributions for a cosmological
model with seven parameters. For illustration purposes we display the average $R$-statistic
 as a function of number of models computed.
The univariate Gaussian proposal 
distribution using fixed variances but adaptive overall step size $\alpha$
(black dashed curve) shows convergence after about 2800 iterations. 
The multivariate Gaussian proposal distribution
 with covariance matrix estimated from the previous chain points and adaptive 
overall step size $\alpha$ as suggested in this
paper (red solid curve) converges after 500 iterations.}
\label{fig::convergence}
\end{center}
\end{figure}

The convergence can be further improved by scaling the covariance
matrix $\bi{S}$ with a variable factor $\alpha$. Using $\alpha$, we
can cope better in situations where the projected likelihood takes on
banana shapes such as in \cite{Kosowsky:2002zt}.  It also improves the
convergence during the early stages when the low number of points
available limits the estimate of the covariance matrix $\covM$.  We dynamically
increase $\alpha$ if a chain takes steps too often,\footnote[2]{Frequent 
acceptance means that the likelihood at the next
  step is roughly comparable to the current one. This happens when the
  chain rarely takes steps larger than $1\sigma$.}  while we decrease
$\alpha$ if the acceptance rate is too low.\footnote{Rare acceptance
  means that the chain rarely explores points with the same
  likelihood, i.e. neighboring points within $1 \sigma$.}
The positive effect of our scheme on the convergence
is illustrated in figure \ref{fig::convergence}.

One can show that modifying the proposal distribution  based on previous chain data during
the run may lead to  a wrong stationary distribution $\pi'(\theta|X) \ $\cite{Neal,Gamerman}.
Therefore, we only apply the  dynamical strategy of finding an optimal step proposal during
the early stages of the simulation. When the convergence
is better than $R=1.2$ and the chain has calculated a certain number of points,
we freeze in  the step proposal distribution. All points in the chain before
this freeze in should be discarded (see also \ref{appendix::format}).

%%%%%%%%%%%%%%%%%%%%%%%%%%%%%%%%%%%%%%%%%%%%%%%%%%%%%%%%%%%%%%%%%%%%%%%%%%%%%%%%%%%%%%%%%%%%%%%%%%

%%%%%%%%%%%%%%%%%%%%%%%%%%%%%%%%%%%%%%%%%%%%%%%%%%%%%%%%%%%%%%%%%%%%%%%%%%%%%%%%%%%%%%%%%%%%%%%%

%%%%%%%%%%%%%%%%%%%%%%%%%%%%%%%%%%%%%%%%%%%%%%%%%%%%%%%%%%%%%%%%%%%%%%%%%%%%%%%%%%%%%

%%%%%%%%%%%%%%%%%%%%%%%%%%%%%%%%%%%%%%%%%%%%%%%%%%%%%%%%%%%%%%%%%%%%%%%%%%%%%%%%%%%%%%%%%%%%%%%%%%%
\section{The Software}\label{sec::user}

The package is  part of {\sc Cmbeasy} and 
consists of two main components. The first one is a
MCMC driver using LAM/MPI \cite{lam} for parallel execution of  each
chain. The second one is the \class{AnalyzeThis} class which is designed to
evaluate the likelihood of a given model with respect to various data
sets. These sets include the latest data of WMAP TT and TE \cite{Hinshaw:2003ex,Kogut:2003et},
ACBAR \cite{Kuo:2002ua}, CBI \cite{Readhead:2004gy}, VSA \cite{Dickinson:2004yr} , 2dFGRS
\cite{Percival:2001hw,Verde:2002,Peacock:2001gs}, SDSS \cite{Tegmark:2003ud,Tegmark:2003uf},
the SNe Ia
compilations of Riess et al. \cite{Riess:2004nr}, Tonry et al.~\cite{Tonry:2003zg} and Knop et. al.
\cite{Knop:2003iy} as well as  the IfA Deep Survey SNe Ia data \cite{Barris:2003dq}.  
Data files for all experiments are included for
convenience. New data sets are added continuously to the code. 

\subsection{The MCMC Driver}
The example MCMC driver consists of two routines: master() and
slave(). Using LAM/MPI for parallel computing, one master and up to
ten slaves may be started. 
The master() will determine the initial random starting position for
each chain. In a never ending  loop, it then sends the parameters
to the slave()'s and collects the results when the computation is finished.
Whenever a step has been successful, it stores the parameters and
likelihoods of the last step together with  the number of  times the chain remained 
at the same point in parameter space in a new line of one file per chain.
The format of these raw MCMC data files is defined in \ref{appendix::format}.
The master() monitors convergence and mixing and determines
the next step for the slave(). Before freeze in, the covariance is 
estimated and the step proposal accordingly modified. After freeze in,
the proposal distribution remains unchanged.

\subsection{The AnalyzeThis Class}\label{analyzethis}
The \class{AnalyzeThis} class provides several routines concerning CMB, SNe Ia
and Large Scale Structure measurements. It also contains routines 
for marginalizing and plotting the Monte Carlo data.

\subsubsection{WMAP}
The WMAP routines are (slightly modified) C++ ports of 
the likelihood code \cite{Verde:2003ey}
available at the LAMBDA \cite{WMAP_Likelihood} web site. 
When
the WMAP routines are called for the first time, the covariance
matrices provided by WMAP will be converted to a binary format to
speed up future use.  A routine for WMAP normalization of the
$C_l$ spectrum using the binned TT may be used instead of the old COBE
normalization. (After the quick normalization, a best fit
  normalization may be called which uses the full likelihood routine
  provided by \cite{Verde:2003ey}.) 

%%%%%%%%%%%%%%%%%%%%%%%%%%%%%%%%%%
\subsubsection{ACBAR, CBI and VSA}
We use the procedures for likelihood computation 
described by the ACBAR, CBI and VSA  collaborations \cite{Kuo:2002ua,Readhead:2004gy,Dickinson:2004yr}, using window
functions and
calibration uncertainty. 
One can de-select data bins from each of these datasets. It is, for instance,
possible to calculate the likelihood for the low $l$ data only, thus speeding
up the computation of a model, because the window functions of the 
high $l$ data  require multipoles up to $l=4000$. One may also wish to exclude $l < 800$ data if one is
using the WMAP observations in order to keep the sets independent.
%%%%%%%%%%%%%%%%%%%%%%%%%%%%%%%%%%
\subsubsection{2dFGRS}
For this Large Scale Structure dataset one compares the data with the theoretical power spectrum 
at $z=0.17$, the effective redshift of the survey,
 multiplied with the window function. We only include the region with $k/h < 0.15 \mbox{ Mpc}^{-1}$ since
at smaller scales nonlinear effects need to be taken into account \cite{Percival:2001hw}. For these values, the 
bias is nearly constant \cite{Verde:2002,Peacock:2001gs}. 
One may either specify the bias or marginalize over it.
%%%%%%%%%%%%%%%%%%%%%%%%%%%%%%%%%%%%
\subsubsection{SDSS}
The theoretical power spectrum to be compared with the data 
should be evaluated at the effective redshift of this survey, at $z=0.1$. 
We use the data given in Table 3 of 
\cite{Tegmark:2003ud} and the appropriate window functions. One may select the maximum
 $k$-value to be included in the likelihood estimate, but including data beyond $k/h > 0.15 \mbox{ Mpc}^{-1}$
requires nonlinear corrections. Again, one can specify the bias or marginalize over it.

%%%%%%%%%%%%%%%%%%%%%%%%%%%%%%%%%%
\subsubsection{Supernovae Ia}
We include four routines for calculating the likelihood with respect to
SNe Ia data. Please note that the sets of Riess et al., Tonry et al. and Knop et al.
are not independent.

\paragraph{Riess et. al.}
One can use the full dataset, subset selection of the ``gold'' set as described in \cite{Riess:2004nr} is 
possible. Likelihood computation as given in this paper.  

\paragraph{Tonry et. al.:}
From the supernovae compilation of Tonry et al. \cite{Tonry:2003zg} one can use the full data set 
of 230 Supernovae or, alternatively, one may use a restricted set of 172 supernovae, where supernovae
with $z <0.01$ and with excess reddening have been omitted as suggested in \cite{Tonry:2003zg}.
 In any case, we have taken the 
particular velocity uncertainty to be $v=500 \mbox{ km/s}$ corresponding to $\Delta z=0.00167$
 and computed the corresponding uncertainty in the luminosity distance to obtain the likelihood:
\begin{eqnarray}
\chi^2 = \sum_{i=1}^{N} \frac{(\log d_l^{exp}-\log d_l^{th})^2}{\sigma_{tot}^2}, \\
\mbox{with } \sigma_{tot}^2= \sigma_{exp}^2+ \left(\frac{d \log d_l^{th}}{d z}\right)^2\Delta z^2.
\end{eqnarray}
Here, $d_l^{exp}$ and $d_l^{th}$ is the experimental and theoretical luminosity distance, respectively.

\paragraph{Knop et. al.:} The 54 SNe Ia presented in Tables 3-5 \cite{Knop:2003iy} have been included in the
code. We use the data with strech and extinction correction applied, subsample selection as
discussed in \cite{Knop:2003iy} is also possible.

\paragraph{IfADS survey:} One may choose to include  all 23 SNe Ia  given in  \cite{Barris:2003dq}, or one may exclude the
 supernovae with excess reddening and those not unambigously identified as SNe Ia as described 
in \cite{Barris:2003dq}.

%%%%%%%%%%%%%%%%%%%%%%%%%%%%%%%%%%%%%%%%%%%%%

\subsection{The Graphical User Interface}
The gui may be used to process the raw output files of the MCMC chains. After
starting cmbeasy, the first step is to ``distill'' the chain data files. By distilling we mean
the merging of the raw chain output files into one file. In addition, distilling removes
all burn-in data.
To get started immediately, we include 
raw data from an example MCMC run  in the resources directory of cmbeasy. The four chains are called montecarlo\_chain$\alpha$.dat.
 They are runs for a $\Lambda$CDM model and need to be distilled first.
Two and one dimensional marginalized likelihoods may then be plotted and printed from within
the gui (see figure \ref{fig::gui}).
Please see the ``howto-montecarlo'' document shipped with cmbeasy for an introduction.

%%%%%%%%%%%%%%%%%%%%%%%%%%%%%%%%%%%%%%%%%%%%%%%%%
\section{Cosmological constraints from the data sets}\label{sec::constraints}
Having described the method and software, we can now proceed to investigate the impact of the different data sets 
on the distribution of parameters for a given cosmological model. For illustrative purposes we limit 
ourselves to a flat $\Lambda$CDM cosmology with five parameters. We take the reduced baryon and matter densities
 $\Omega_b h^2$ and $\Omega_m h^2$, the Hubble parameter $ h=H_0/(100 \mbox{km s}^{-1} \mbox{Mpc}^{-1} )$, the optical 
depth to the last scattering surface $\tau$ and the spectral index of the initial
 power spectrum $n_s$ as parameters.
 We neglect
any tensor contributions and we marginalize over the amplitude of the initial power spectrum. Thus, the amplitude is 
treated as a ``nuisance'' parameter that is integrated out. 

 In table \ref{priors} we display the cosmological
 parameters and the flat priors used. 
%%%%%%%%%%%%%%%%%%%%%%%%%%%%%%%%
\begin{table}
\caption{\label{priors} Flat priors on the parameters used in our MCMC simulations}
\begin{indented}
\item[]\begin{tabular}{@{}ccc}
\br
Parameter & Min  & Max \\
\mr
$\Omega_b h^2$ & 0.016  & 0.03 \\
$\Omega_m h^2$ & 0.05 &  0.3 \\
$h$ & 0.60  & 0.85 \\
$\tau$ & 0 &  0.9 \\
$n_s$ & 0.8 & 1.2\\
\br
\end{tabular}
\end{indented}
\end{table}
%%%%%%%%%%%%%%%%%%%%%%%%%%%%%%%%%%%%%%%%%%%%%%%%%%%%%%%%%%%%%%%%%%%%%%%%%%%%%%%%%%%%%%%
The limits of the parameter ranges are chosen in accordance with previous results 
\cite{Bennett:2003bz,Tegmark:2003ud}.

First, we will investigate constraints from CMB-only data, subsequently adding Supernovae Ia and large scale 
structure data sets. For a detailed description of how we implemented the datasets we refer the reader to 
section \ref{analyzethis} of this paper. We have performed three MCMC simulations with CMB only, 
CMB+ SNe Ia and CMB + SNe Ia + LSS data sets.  Convergence was reached in each
run after about $\sim 2000$ iterations, each simulation was run until 55 000 models had been computed.

\begin{figure}[!t]
\begin{center}
\includegraphics[angle=0,scale=0.5]{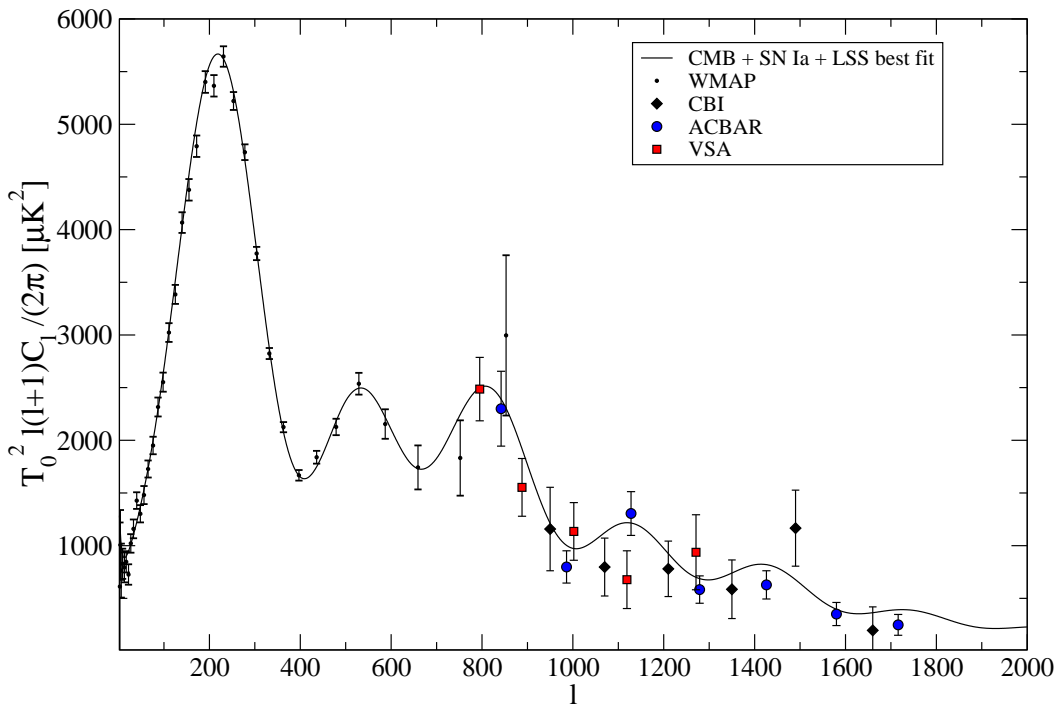}
\caption{CMB data set for constraining cosmological parameters.}
\label{fig::cmb}
\end{center}
\end{figure}

\begin{figure}[!t]
\begin{center}
\includegraphics[angle=0,scale=0.5]{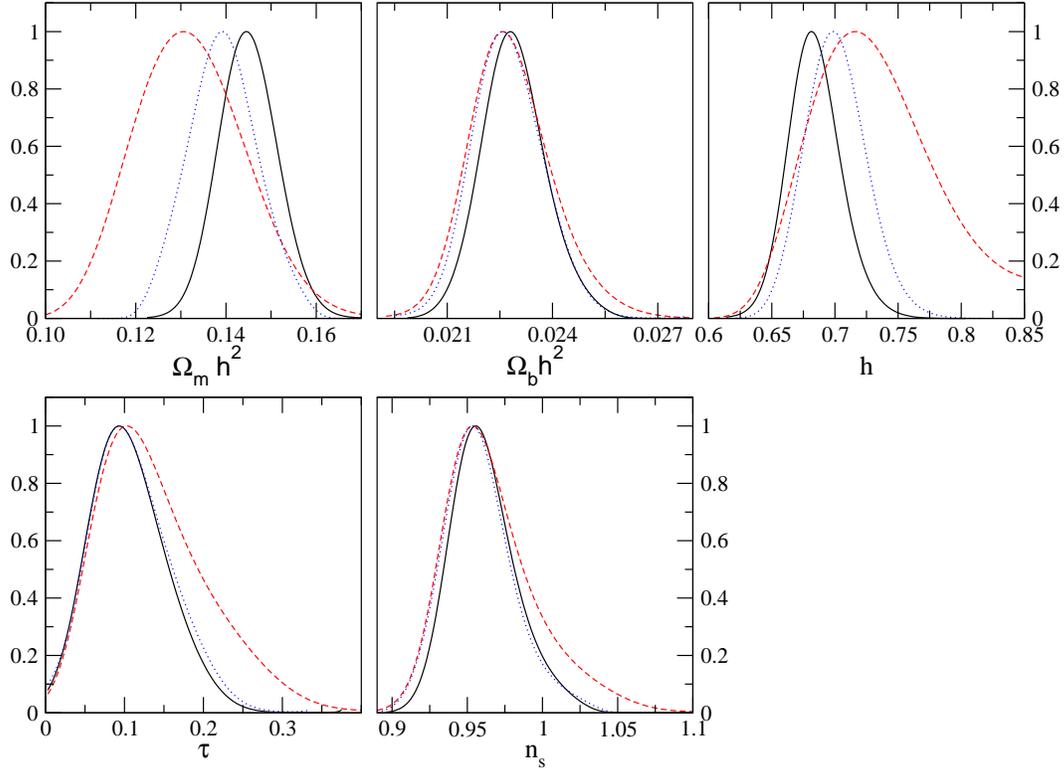}
\caption{One dimensional marginalized distributions for the parameters. Constraints from CMB only (red, dashed), 
CMB + SNe Ia (blue, dotted) and CMB + SNe Ia + LSS (black, solid). The histograms extracted from the chains were 
fitted using a function $f(x)=\exp(p(x))$ with $p(x)$ a sixth-order polynomial \cite{Tegmark:2003ud}. }
\label{fig::oned}
\end{center}
\end{figure}

For the CMB-only run we use the measurements of WMAP (TT and TE spectra), CBI, VSA and ACBAR up to $l=2000$, 
removing data points from CBI, VSA and ACBAR
 where we have WMAP measurements with comparable  error bars to keep the data sets independent.
 The entire set we used is displayed in 
figure \ref{fig::cmb}. 
%%%%%%%%%%%%%%%%%%%%%%%%%%%%%%%%%%%%%%%%%%%%%%%%
%%%%%%%%%%%%%%%%%%%%%%%%%%%%%%%%%%%%%%%%%%%%%%%% 
For CMB + SNe Ia we add the data set of Riess et al. \cite{Riess:2004nr} (``gold sample''), and  finally for  CMB + SNe Ia + LSS
we add the large scale structure measurements of the SDSS collaboration using all points with
 $k/h < 0.15\; \mbox{Mpc}^{-1}$ where the perturbations 
are still linear \cite{Tegmark:2003uf}. 
 The results are presented in table \ref{tab::constraints}, marginalized likelihoods
for the parameters are displayed in table \ref{fig::twod} and figure \ref{fig::oned}.
%%%%%%%%%%%%%%%%%%%%%%%%%%%%%%%%%%%%%%%%%%%%%%%%
%%%%%%%%%%%%%%%%%%%%%%%%%%%%%%%%%%%%%%%%%%%%%%%% 

%%%%%%%%%%%%%%%%%%%%%%%%%%%%%%%%%%%%%%%%%%%%%%%%%
\begin{table}
\caption{\label{tab::constraints}Constraints on the parameters from a combination of data sets. These confidence intervals
were generated from the one-dimensional marginalized distributions
 (errors are given at 68.3 \% confidence level).}
\begin{indented}
\renewcommand{\arraystretch}{2}
\item[]\begin{tabular}{@{}cccc}
\br
Parameter & CMB only  & CMB + SNe Ia & CMB+ SNe Ia + LSS \\
\mr
$\Omega_b h^2$ & $0.02261^{+0.0012}_{-0.0011}$  & $0.02257^{+0.0011}_{-0.0009}$ &$0.02278^{+0.0009}_{-0.0009}$ \\
$\Omega_m h^2$ & $0.1306^{+0.013}_{-0.012}$ &  $0.1393^{+0.0076}_{-0.008}$ &$ 0.144^{+0.0063}_{-0.0066}$ \\
$h$ & $0.716^{+0.053}_{-0.041}$  & $0.699^{+0.024}_{-0.024} $ & $0.682^{+0.021}_{-0.021}$ \\
$\tau$ & $0.1019^{+0.081}_{-0.52}$ & $0.0941^{+0.058}_{-0.043}$ &$0.0938^{+0.051}_{-0.0043}$\\
$n_s$ & $0.954^{+0.032}_{-0.024}$ & $0.953^{+0.024}_{-0.021}$ & $0.956^{+0.023}_{-0.019}$\\
\br
\end{tabular}
\end{indented}
\end{table}
%%%%%%%%%%%%%%%%%%%%%%%%%%%%%%%%%%%%%%%%%%%%%%%%%%%%%%
%%%%%%%%%%%%%%%%%%%%%%%%%%%%%%%%%%%%%%%%%%%%%%%%%%%%%

\begin{table}
\caption{\label{fig::twod}Two-dimensional marginalized likelihoods for the cosmological parameters 
using CMB +  SNe Ia + LSS data. The contours are one, two and three sigma confidence 
regions, respectively.}
\begin{tabular}{cccc}
\setlength{\tabcolsep}{0mm}
\includegraphics[angle=0,scale=0.18]{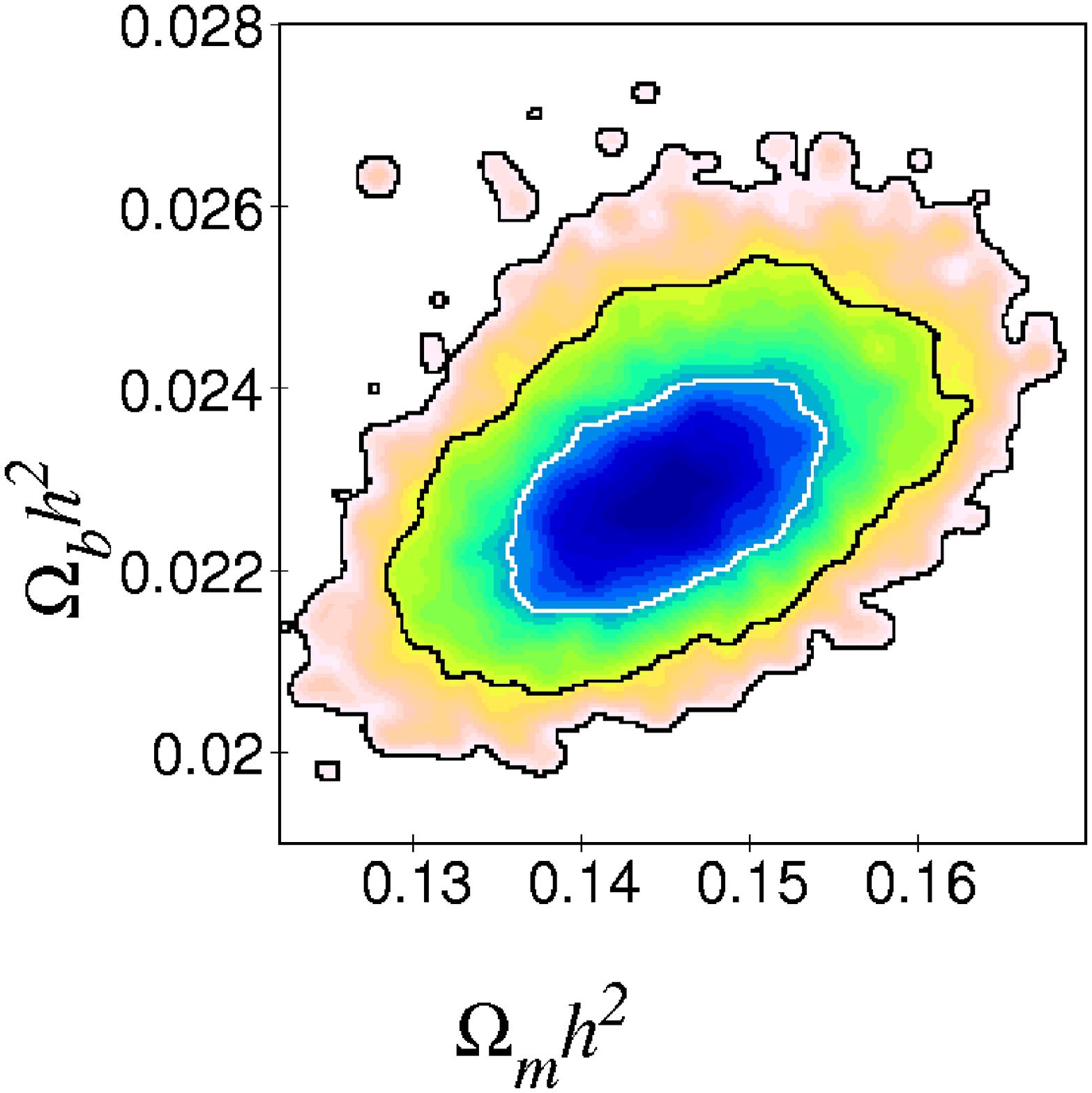} &\includegraphics[angle=0,scale=0.18]{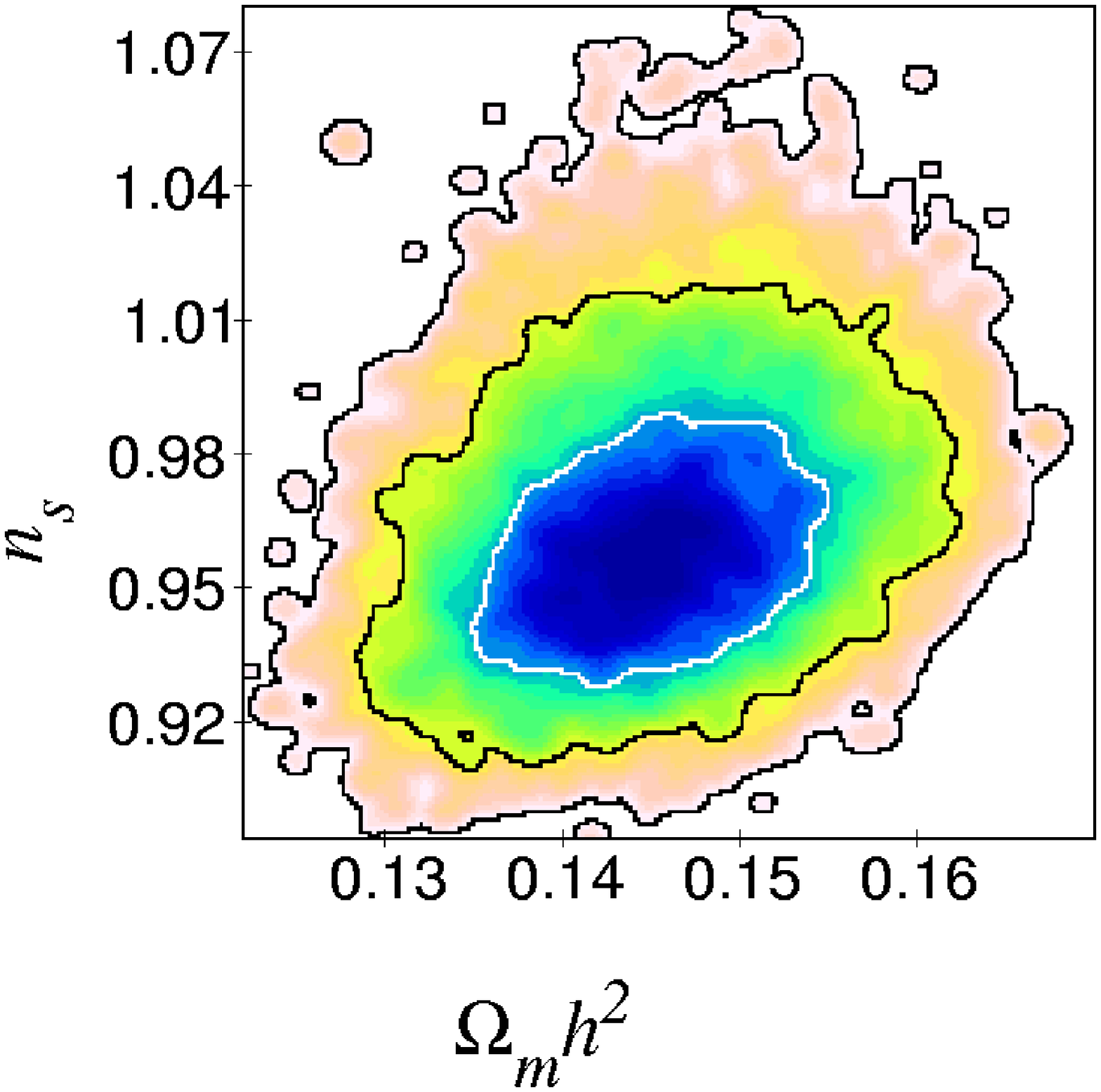}   & 
\includegraphics[angle=0,scale=0.18]{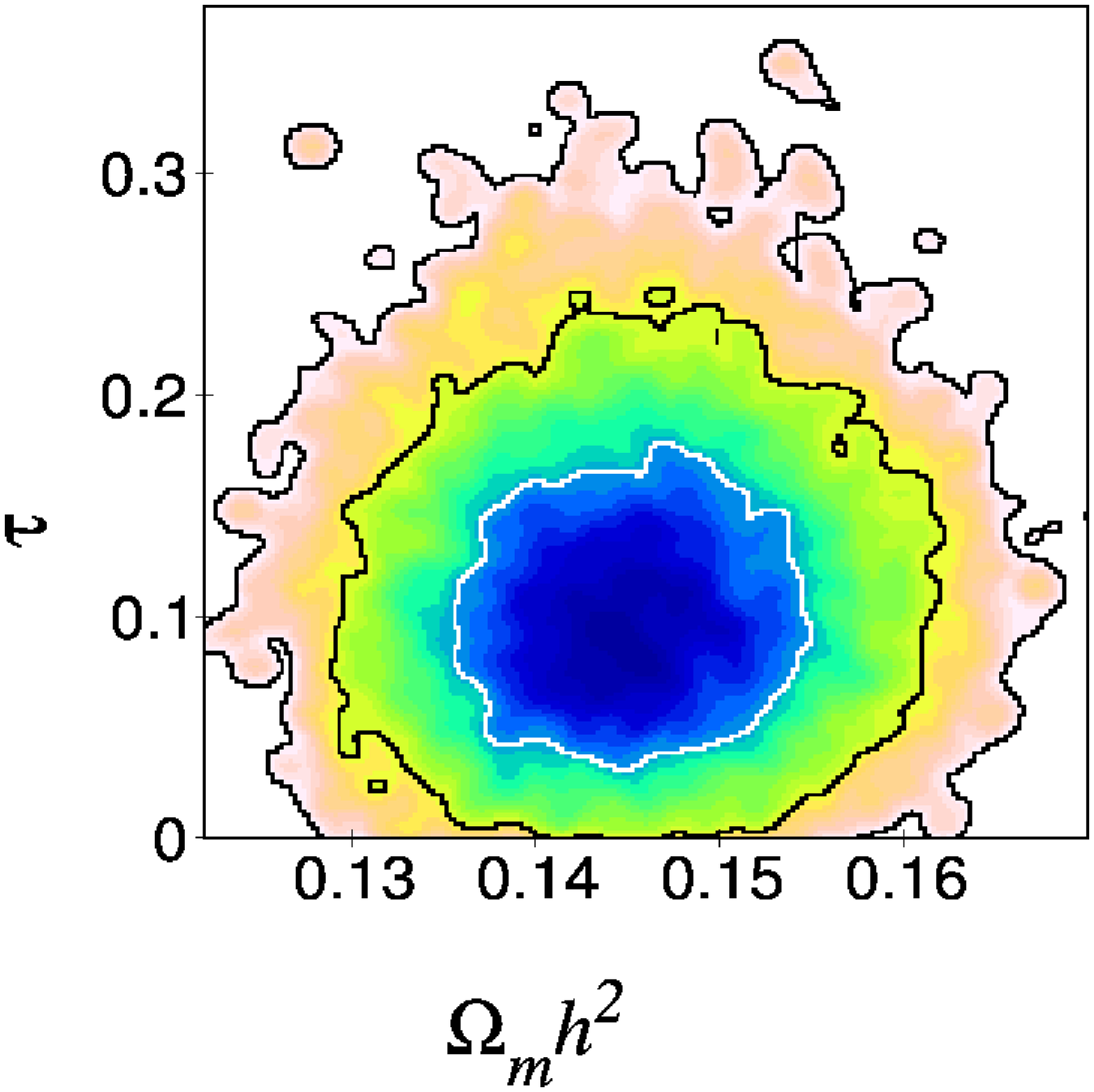} &\includegraphics[angle=0,scale=0.18]{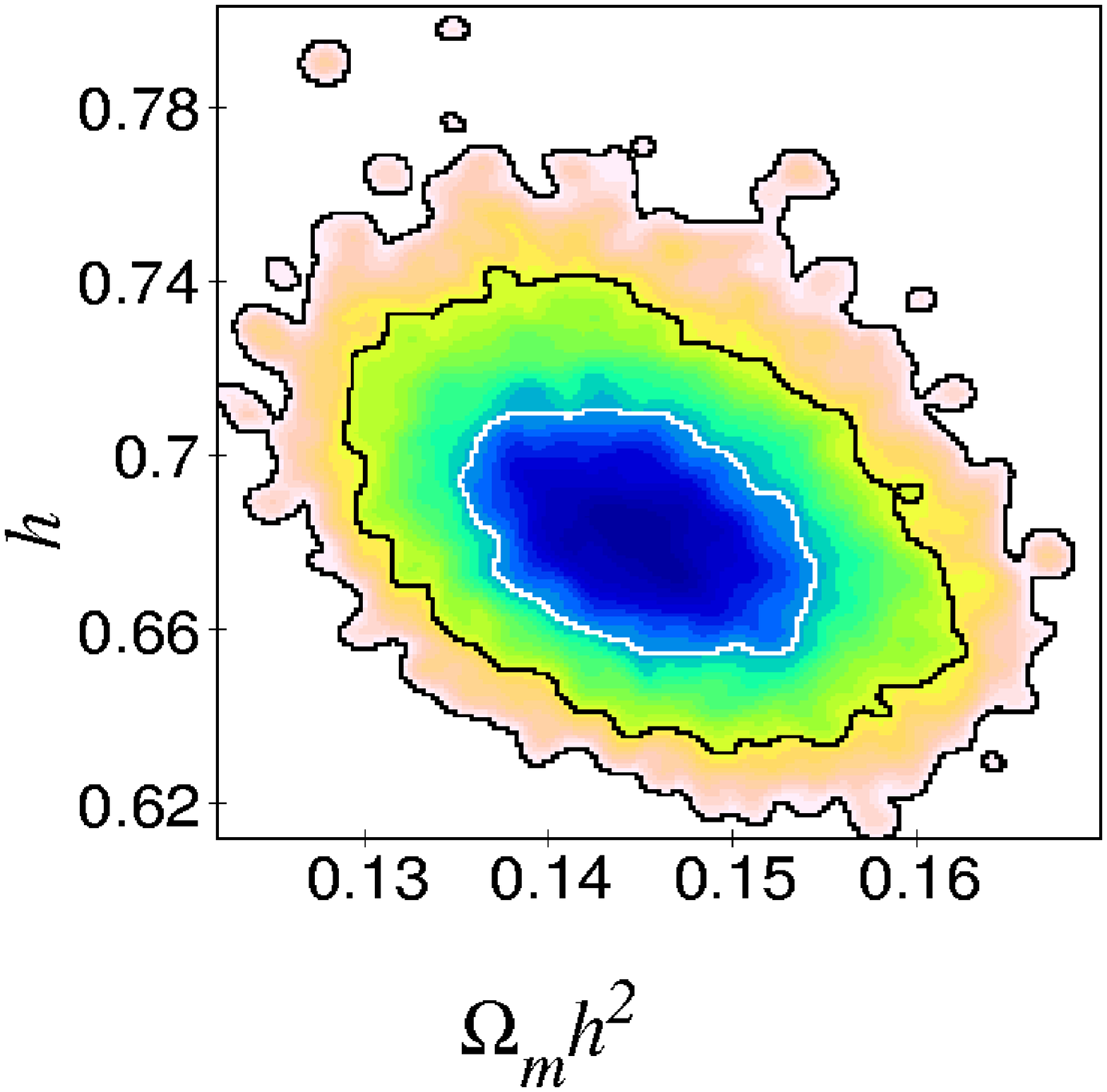} \\
\includegraphics[angle=0,scale=0.18]{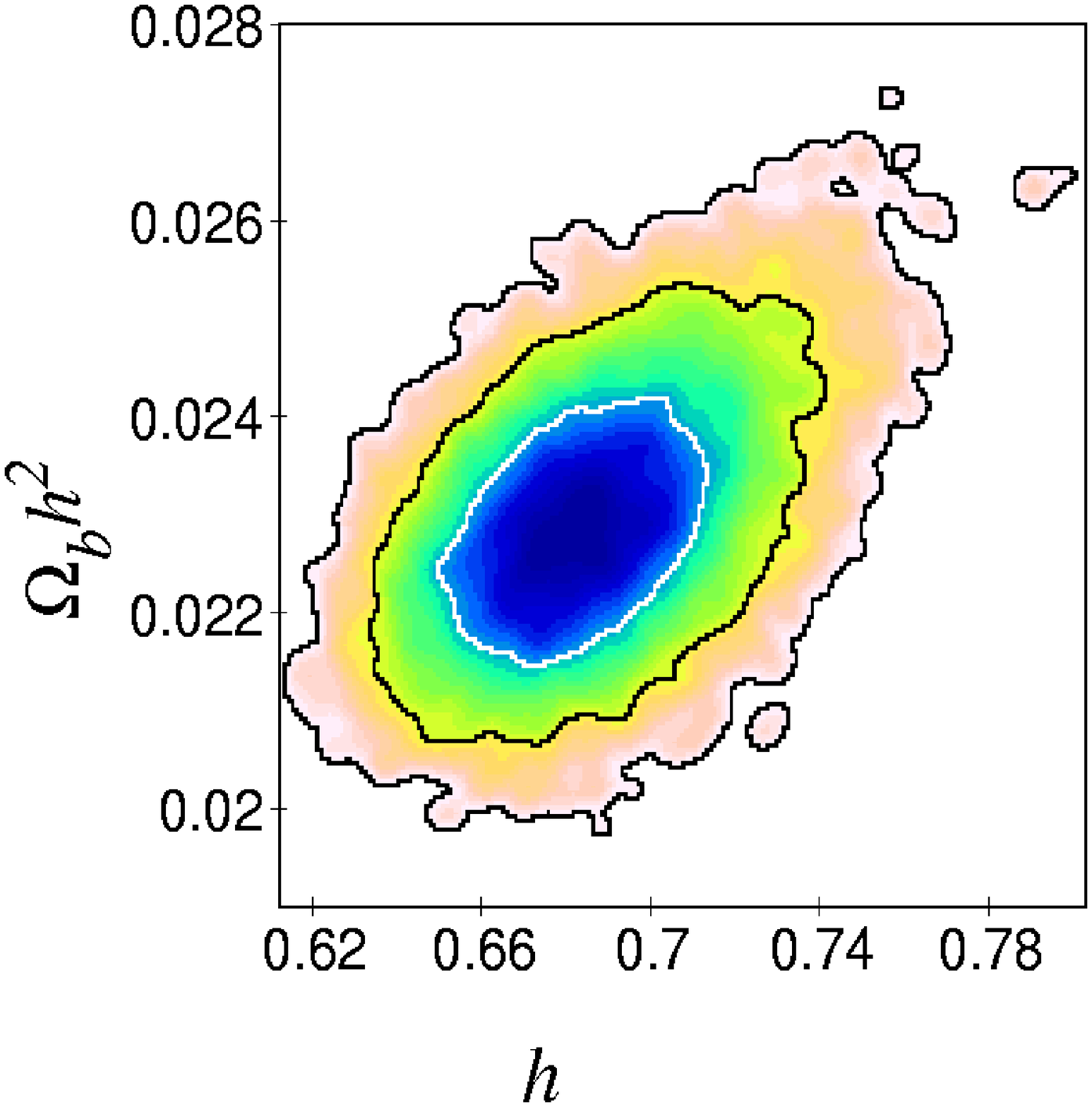}& \includegraphics[angle=0,scale=0.18]{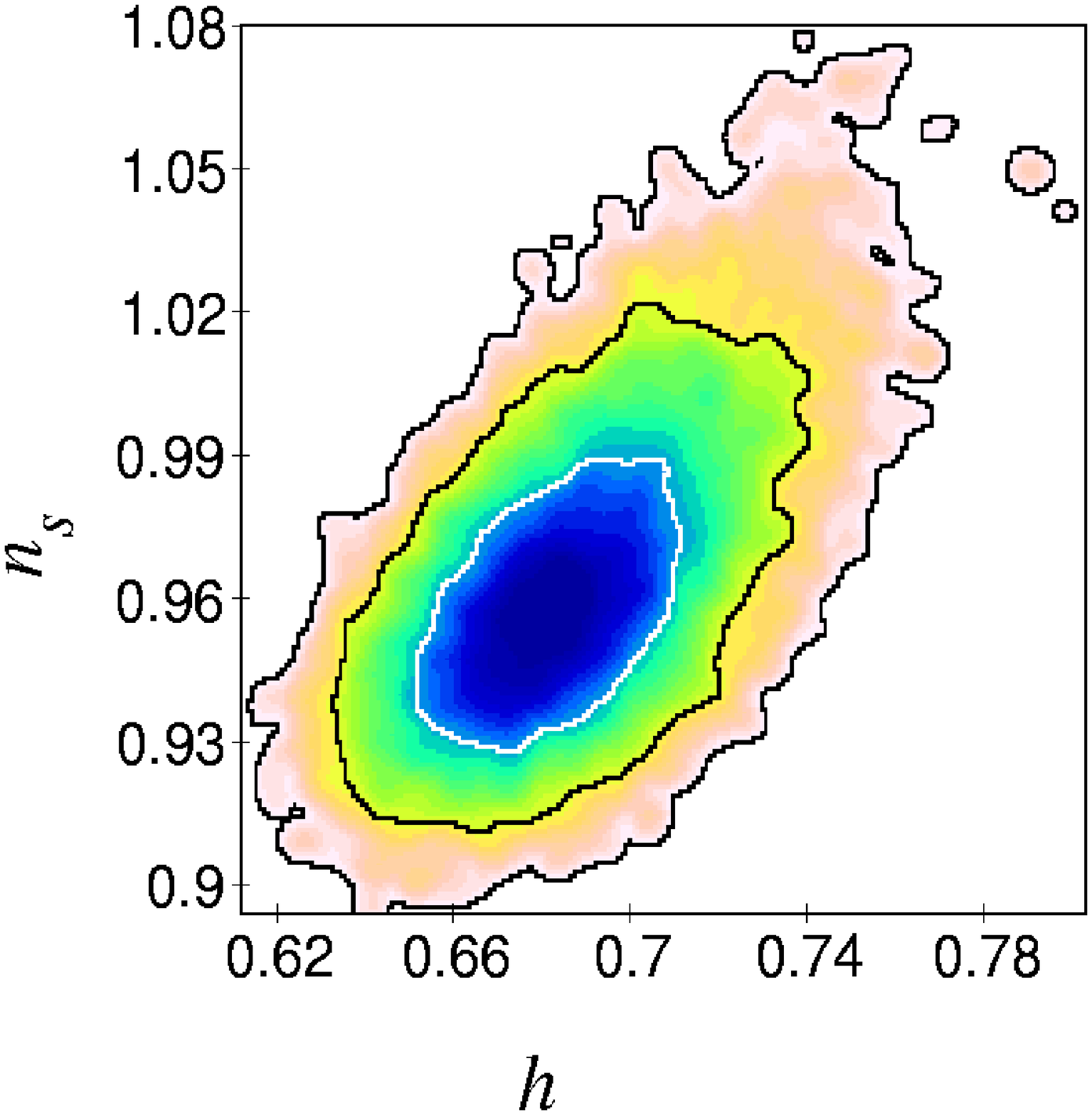} &
 \includegraphics[angle=0,scale=0.18]{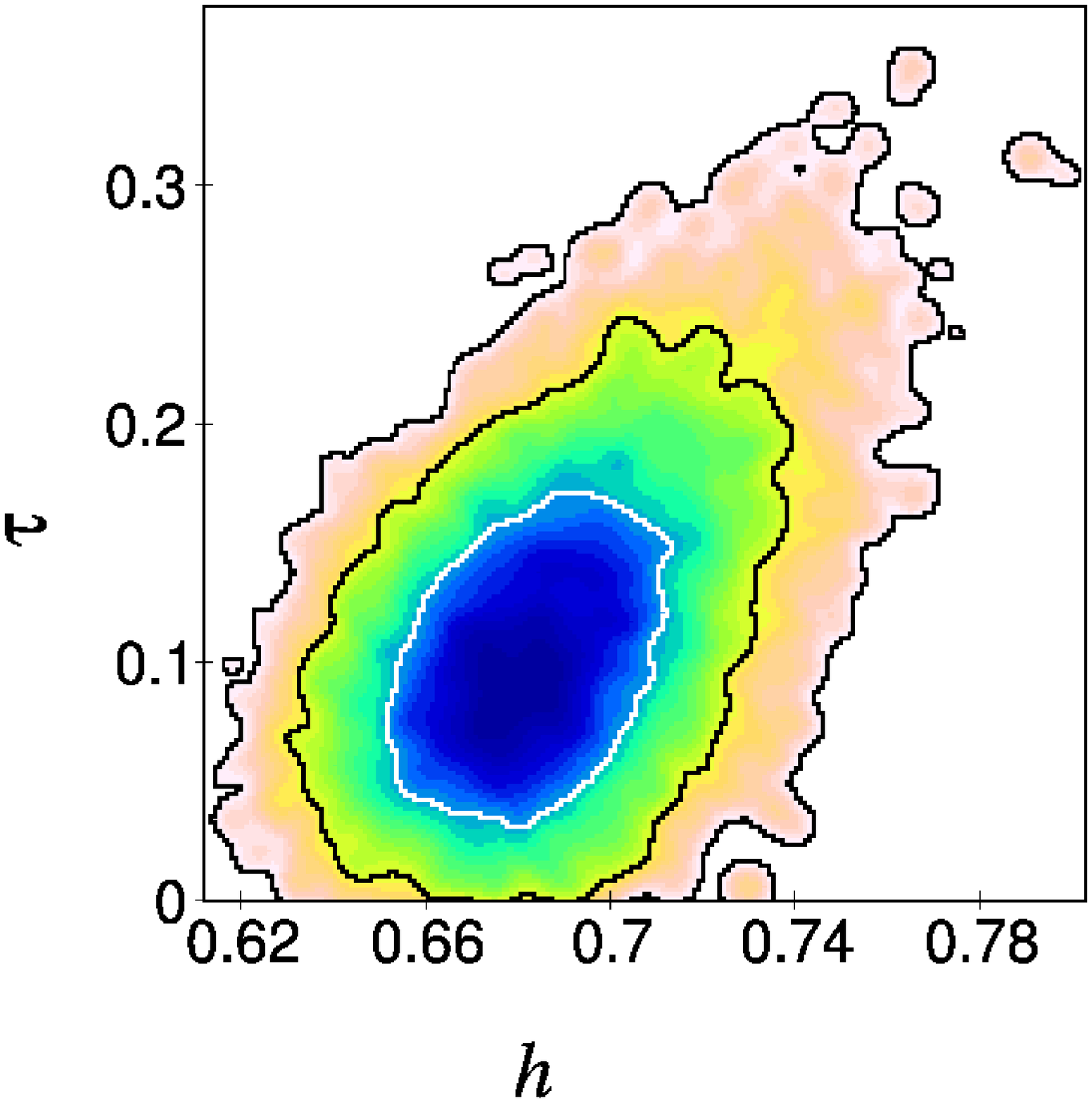} &  \\
\includegraphics[angle=0,scale=0.18]{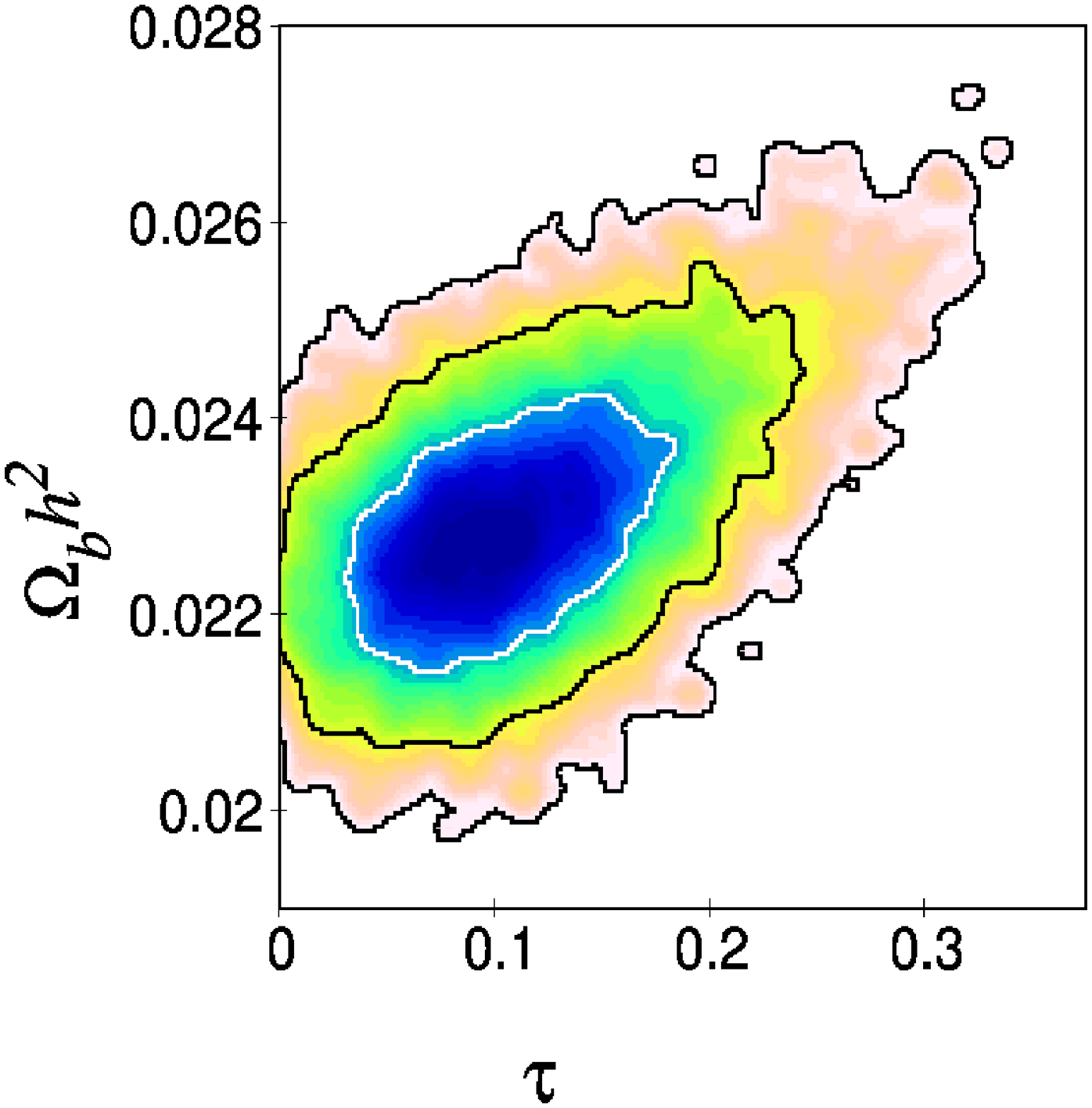} & \includegraphics[angle=0,scale=0.18]{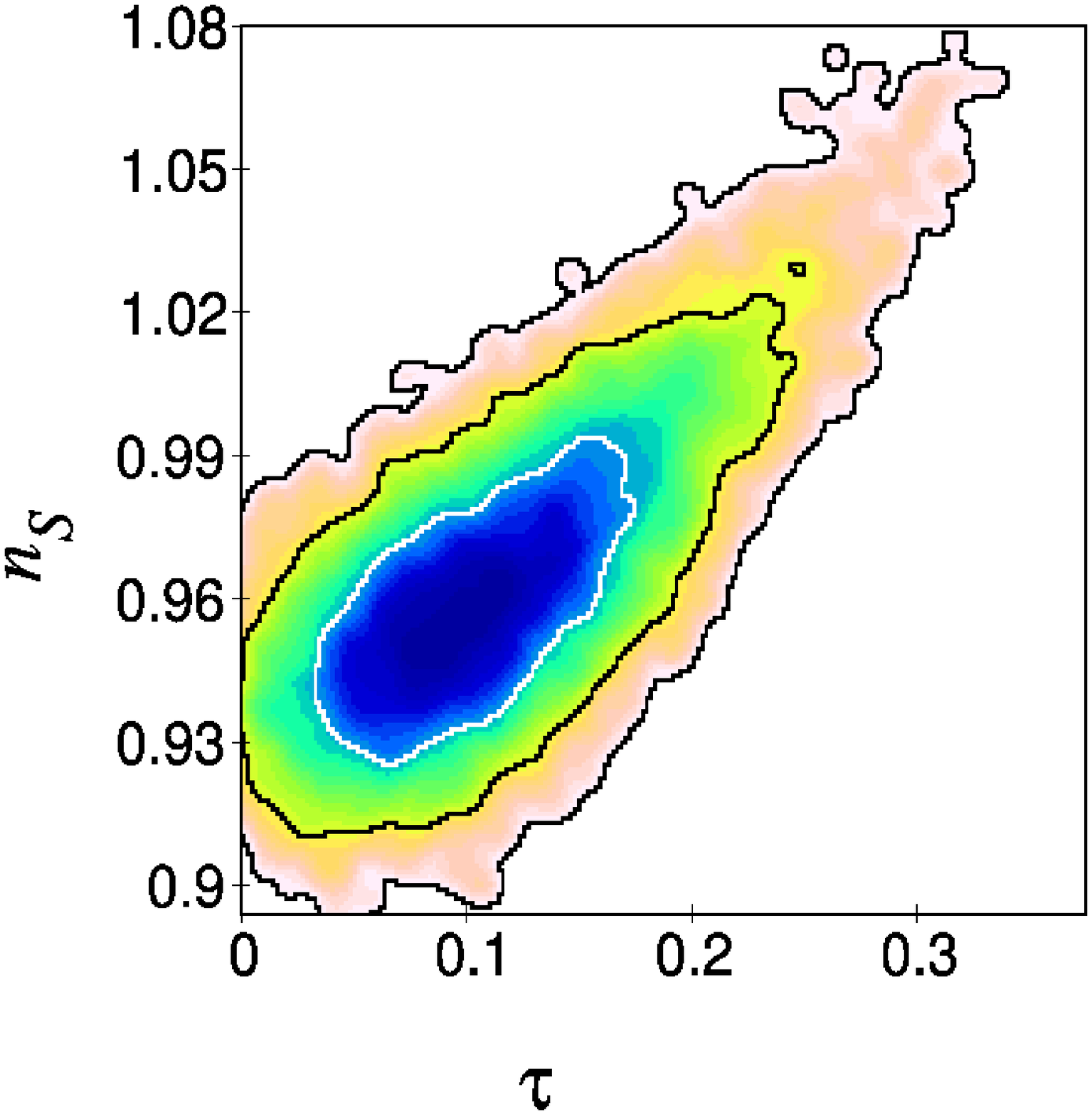}&  & \\
\includegraphics[angle=0,scale=0.18]{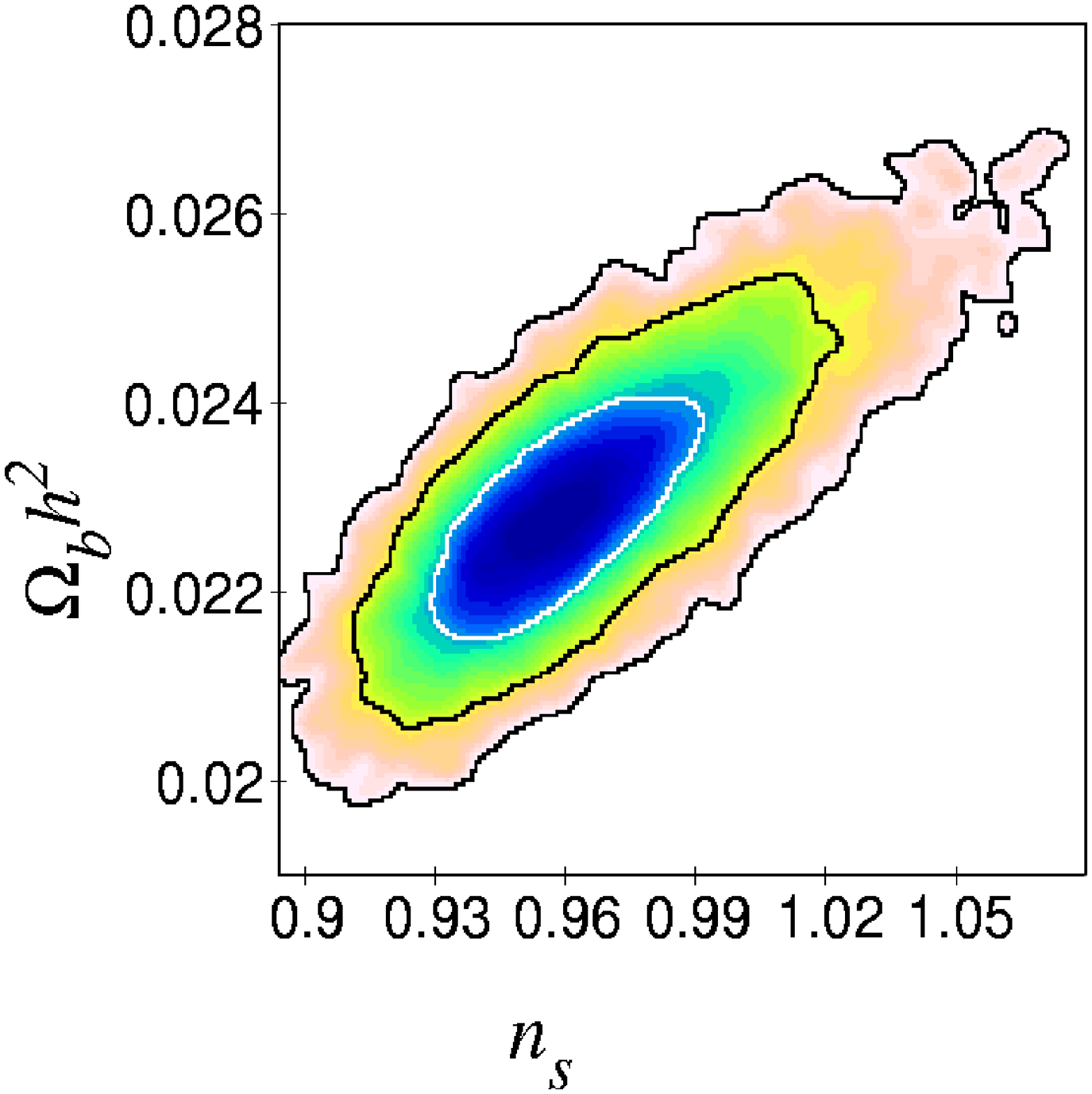} & & & \\

\end{tabular}
\end{table}

%%%%%%%%%%%%%%%%%%%%%%%%%%%%%%%%%%%%%%%%%%%%%%%%%%%
We will now discuss the impact of the different sets on the parameters. 
 Consider figure \ref{fig::oned}. The CMB-only data set does not well constrain $\Omega_m h^2$. Adding supernovae improves the bounds on the total matter content considerably, because the lumiosity
distance depends sensitively on $\Omega_m$.
 Adding large scale structure data breaks the  degeneracy between matter contribution and initial power spectrum amplitude, 
hence we obtain still tighter bounds. Note
that the distribution of  $\Omega_m h^2$ obtained from CMB-only and CMB + SNe Ia + LSS data sets differ somewhat. 
This has already been noted in \cite{Tegmark:2003ud}. Even from CMB measurements alone we can infer the presence of dark matter
at high significance.  
In contrast, the baryon contribution $\Omega_b h^2$ is already well constrained by the CMB alone, adding SNe Ia and LSS data only
 improves the
bounds slightly. The CMB is most sensitive to this parameter since what one observes are essentially oscillations in a photon-baryon 
plasma, the density of baryons is a critical parameter for the shape of the CMB spectrum. SNe Ia and LSS measurements
are more or less insensitive to this parameter, though hints at oscillations in the power spectrum have 
been detected \cite{Percival:2001hw}. 

The Hubble parameter $h$, is only slightly constrained
by CMB measurements alone. Adding SNe Ia data considerably improves the bounds, adding large scale structure data does not improve the
bounds very much. The bounds obtained for the Hubble parameter for the full data set 
 are consistent with the HST key project value $h=0.72\pm 0.08$
\cite{Freedman:2000cf} derived from measuring the local Hubble flow. 
Even though the optical depth does not directly influence SNe Ia predictions, adding SNe Ia data tightens
the bound on the optical depth $\tau$ as seen in figure \ref{fig::oned}. The mechanism is
somewhat indirect: SNe Ia data tightens the bound on $\Omega_m$ and thus limits
the range of allowed values for $\tau$, $h$ and $n_s$ from CMB measurements. As we marginalize
over the bias of the SDSS data, large scale structure does not add further to the bounds on $\tau$.

%%%%%%%%%%%%%%%%%%%%%%%%%%%%%%%%%%%%%%%%%%%%%%%%%%%%%%%%%%%%%%%%
\section{Conclusions}\label{sec::conclusion}

We have introduced the {\sc AnalyzeThis} package, which can be used
to constrain cosmological models using observational data sets.
The \class{AnalyzeThis}  class provides many functions to compute the likelihood of a model with respect
to  measurements of the CMB, SNe Ia and Large Scale Structure.\footnote[8]{We would be happy to include
contributions of routines for recent and future measurements. So if the readers favorite measurement
is not included yet, please send in a few lines of code.}

In order to constrain models of the Universe with a substantial number
of parameters, we include a Markov Chain Monte Carlo driver. 
As the MCMC step strategy 
determines the convergence speed of the chains, we implemented a multivariate Gaussian sampler
with an additional dynamical scaling. We stop the  adaptive improvement of the step proposal density 
as soon as a good level of convergence is reached and discard all data calculated before. Using
this approach, we combine the advantage of a single-run automatic optimization of the step strategy
with the demand of a static step proposal density. 

The output of the Monte Carlo chains is in a human readable format and may be processed by 
any software, even during the run. For convenience however, 
one may use {\sc Cmbeasy}'s  gui to marginalize, 
plot and print two and one dimensional likelihood surfaces.

Finally, we discussed the impact of the different data sets on the parameters for the case of a $\Lambda$CDM cosmology
with five parameters.

%%%%%%%%%%%%%%%%%%%%%%%%%%%%%%%%%%%%%%%%%%%%%%%%%%%%%%%%%%%%%%%%%%%%%%%%%%%%%%%%%%%%%%%%%%%%%%%%%%%%%
\ack
We would like to thank Robert R.~Caldwell, Robert A.~Knop,  Havard B.~Sandvik, Max Tegmark, Adam G.~Riess and Timothy J.~Pearson
 for helpful discussions. 
M. Doran is supported by NSF grant PHY-0099543.  C.M. M{\"{u}}ller is supported by GRK grant 216/3-02.

%%%%%%%%%%%%%%%%%%%%%%%%%%%%%%%%%%%%%%%%%%%%%%%%%%%%%%%%%%%%%%%%%%%%%%%%%%%%%%%%%%%%%%%%%%%%%%%%%%%%%% 

%%%%%%%%%%%%%%%%%%%%%%%%%%%%%%%%%%%%%%%%%%%%%%%%%%%%%%%%%%%%%%%%%%%%%%%%%%%%%%%%%%%%%%%%%
\appendix
\section{Files and Formats}\label{appendix::format}
\subsection{MCMC chain output data}
The parameters and likelihoods of each model are stored in files  with the naming
convention ``montecarlo\_chain$\bi{\alpha}$.dat'', where ``$\bi{\alpha}$'' is an integer.
Each file corresponds to a slave() process (and hence to an independent chain). 
Each line in such a file represents a new point in parameter space. The format of 
one such line is:\\
\[
p_0\ |\ p_1\ | \dots |p_n\ | \  L \ | \ l_0 \ |  \ \dots |\ l_m \ | \  a_1 \ | \ \dots | \ a_k \ | \ M,
\]
where $p_i$ are the parameters of the model, $L$ is the overall likelihood, the $l_i$
are likelihoods of different experiments, $a_i$ are auxiliary fields (to store $\sigma_8$,
for instance). Finally, $M$ determines the weight of this parameter set, i.e. the ``time''
spend before leaving this point in parameter space. 
When the step sizes and the covariance estimates are frozen in, one line in each
file gets a zero $M$. All data before this line should be regarded as ``burn-in''.
The graphical user interface (using the distillChains() routine of \class{AnalyzeThis}) will
automatically discard all models before freeze-in.

\subsection{Monitoring Progress and Covariance}
The master() routine outputs the convergence and some more information into
a file called ``progress.txt''. Until the step proposal distribution is frozen in,
the covariance matrix is in addition output to ``covMatrix.txt''. 
Messages of any errors occuring in slave() will be appended to the file ``errorlog.txt''.
 
\vspace{1cm}

\bibliographystyle{unsrt}

\end{document}